# Theory and mitigation of motional eddy current in high-field eddy current shielding

Seung-Kyun Lee*, Yihe Hua

GE HealthCare Technology & Innovation Center, Niskayuna, NY 12309, USA

**Abstract**

Eddy current shielding by a Faraday cage is an effective way to shield alternating-current (AC) magnetic fields in scientific instrumentation. In a strong static magnetic field, however, the eddy current in the conductive shield is subject to the Lorentz force which causes the shield to vibrate. In addition to mechanical issues (e.g., acoustic noise), such vibration induces motional eddy current in the shield that can dominate the original, electromagnetic eddy current to undermine the conductor's shielding capability. In this work we investigate a method to control motional eddy current by making cut-out patterns in the conductor that follow the electromagnetic eddy current image. This effectively limits the surface current of the plate to a single mode, and prevents proliferation of uncontrolled motion-induced surface currents that disrupts eddy current shielding. After developing a comprehensive theory of magneto-mechanical interaction in a conductive plate, the proposed method was tested on a flat-geometry testbed experiment inside a 3 Tesla magnetic resonance imaging (MRI) magnet. It was found that the magnetic field generated by the motional eddy current was much more localized in space and frequency for a patterned copper shield compared to a solid copper. The magnetic field of the patterned shield could be accurately predicted from the impedance measurement in the magnet. Implications of our results for improved shielding of gradient fields in high-field MRI are discussed.

*Corresponding author*

Email: lsk@gehealthcare.com

## 1. Introduction

Eddy current shielding of time-varying magnetic fields by conductive plates and shells is widely used in scientific instruments as well as everyday electrical devices. The shielding is based on the fact that at sufficiently high frequencies, all conductors develop eddy current to counteract applied magnetic flux changes, effectively preventing time-varying magnetic fields from entering their interior. When a conductive shield is placed inside a strong static magnetic field, as in a modern high-field magnetic resonance imaging (MRI) scanner, the eddy current is subject to Lorentz force, leading to vibration of the shield. The moving conductor now experiences two kinds of time-varying magnetic flux in its frame of reference: (i) the original, externally applied AC magnetic field, and (ii) the changing flux as it cuts through the flux lines of the static magnetic field[1]. The latter flux induces its own eddy current, sometimes called motional eddy current[2]. Depending on the static field and the vibration amplitude, the motional eddy current can dominate the original, electromagnetic eddy current. In this case the conductor can no longer provide shielding because the field generated by the motional eddy current can be very different in shape and larger in magnitude than the applied field.

The failure of eddy current shielding in strong static magnetic field has long been noted in MRI engineering as the time-varying fields generated by the gradient coils in the bore of a cylindrical magnet were found to heat the superconducting coils despite multiple conductive shells in between them[3][4][5,6]. The heating cannot easily be explained without motional eddy current. It should be noted that most modern MRI gradient coils employ self-shielded design where each field-generating (primary) coil unit is paired with a shielding coil on the outside of and connected in series with the primary coil[7]. This greatly reduces the field reaching the magnet, but the residual "leakage" field can still be substantial, on the order of several milliTesla. This is sufficient to heat the superconducting magnet absent some degree of passive shielding provided by conductive cylinders in the cryostat.

Given the importance of thermal protection of superconducting coils, much effort has been directed to reducing the motional eddy current, through gradient design[3], additional passive shielding[8], warm bore engineering[9][10], and magnet reinforcement[11]. While promising, most structural methods face an important challenge, that is limited space between the driving (gradient) coils and the protected structure (magnet). In high-field compact MRI scanners[12], in particular, any radial space between the gradient and magnet coils comes at a steep cost in terms of manufacturing, performance and energy consumption/efficiency[13].

The purpose of this paper is to investigate a new method to reduce and control the motional eddy current, not through purely structural intervention, but by limiting eddy current modes in the conductor through cut-out patterns. The pattern is defined by the ideal eddy current in the high frequency limit in the absence of vibration. By allowing such eddy current to flow in the conductor, while blocking all other modes perpendicular to this mode, the conductor can support non-motional eddy current that is necessary for shielding[14], while being less prone to unwanted eddy current in the presence of vibration. We first present a comprehensive theory of magneto-mechanical coupling in the dynamics of a conductive plate in a static field. We then demonstrate the proposed method in a flat-geometry testbed experiment at 3T.

## 2. Theory

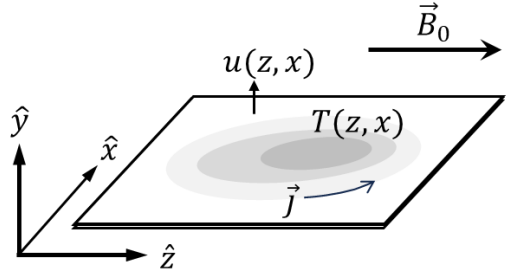

Figure 1. Definition of the coordinate system and variables. A static magnetic field $\vec{B}_0 = B_0 \hat{z}$ is applied parallel to a conductive shield plate in the zx plane whose vibration is characterized by the vertical displacement $u(z,x)\hat{y}$. The eddy current stream function and the corresponding surface current density are denoted by $T$ and $\vec{j}$.

We consider a thin conductive plate with surface conductivity $\sigma_s$ (Fig. 1) that is horizontally placed on the $zx$ plane, and whose vibration consists of vertical (y-directional) displacement $u(z,x)$ that is time-dependent. A static magnetic field $B_0$ is applied in the z direction. Surface eddy current induced in the conductor is characterized by a stream function $T(z,x)$, which is related[15] to the surface current $\vec{j}(z,x)$ by $(j_z, j_x) = (\partial T/\partial x, -\partial T/\partial z) = \nabla \times (T\hat{y})$. Choice of a horizontal zx plane is motivated by the typical coordinate system of a horizontal-bore MRI magnet. The conductive plate is subject to time-varying magnetic field $B_{app}(z,x)\,\hat{y}$ which induces eddy current. If the eddy current completely nulls $B_{app}$ on the surface of the plate, the plate functions as a perfect shield. This will be the case if the plate superconducts, or if the frequency is infinitely high, provided there is no vibration. We will call such an eddy current an eddy image current, whose stream function is denoted as $T_{eddyi}$ (Fig. 2a). At a finite frequency and finite $\sigma_s$, the eddy current will deviate from $T_{eddyi}$. As long as $B_0 = 0$, however, a plate made of copper or aluminum with mm-range thickness and tens of cm-range lateral dimensions will still provide substantial shielding of magnetic fields above a few hundred Hz, a frequency range that is important for thermal protection of MRI magnets.

In the presence of strong $B_0$, on the other hand, the surface current can drastically differ from $T_{eddyi}$, dominated by motional eddy current. This results in passive shielding failure; in fact, a conductive plate can amplify the applied field. In MRI, this phenomenon has been linked to excessive helium boil-off in the magnet cryostat induced by switched gradient fields[3,16,2,1]. The goal of this section is to develop a physical theory to calculate the *total* eddy current in response to the applied field $B_{app}$ that includes the motional eddy current. We will primarily work in the frequency domain at angular frequency $\omega$.

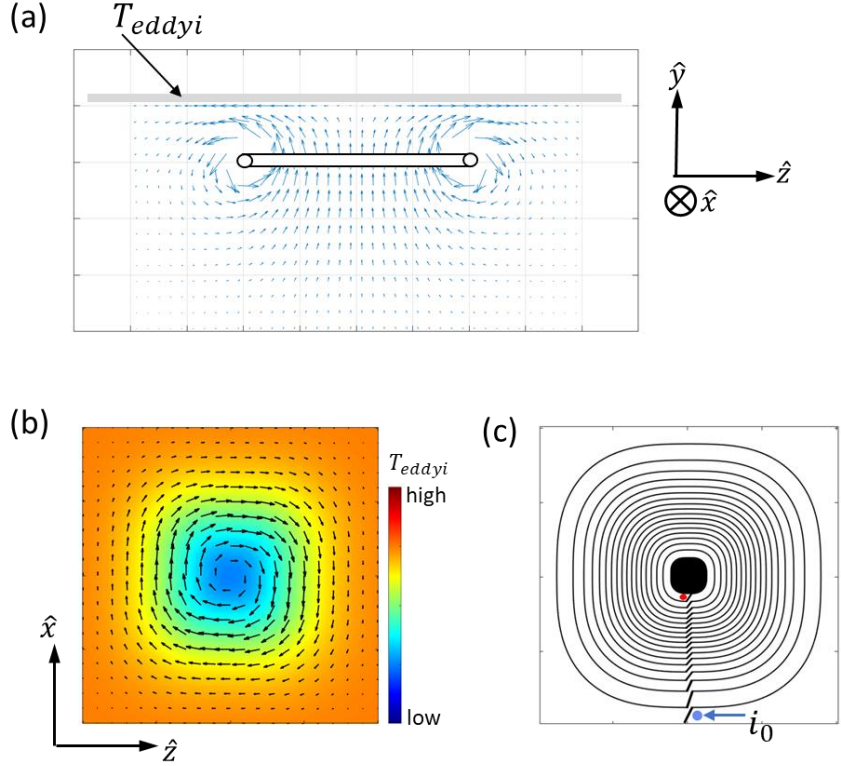

Figure 2. Eddy image current and its discretization. (a) In the high-frequency limit, the eddy current in a conductive plate (gray horizontal bar) prevents penetration of an applied magnetic field generated by a driving coil (black-outlined horizontal bar). Such eddy current, whose stream function is denoted as $T_{eddyi}$, is analogous to the DC screening current in a superconducting plate. (b) Illustration of $T_{eddyi}$ and corresponding surface current map. (c) In the proposed passive shield, $T_{eddyi}$ is discretized into a closed-loop coil with current $i_0$. The black lines indicate cutouts on the conductor and current flows from the red dot in the middle to the blue dot at the bottom (arrow). The two dots are shunted by a copper bridge (not shown) to close the loop and cancel the turn-to-turn radial current.

## 2.1. Magneto-mechanical coupled equations for a conductive plate in a static magnetic field

Detailed derivation of the coupled equations, namely the equation of motion and the circuit equation, is given in Appendix A. Below we describe the equations and discuss their properties. We will use script-font characters $\mathcal{A}, \mathcal{B}, \cdots$ to denote linear operators acting on scalar fields defined in the zx plane. The equation of motion of the plate under the influence of the Lorentz force can be written as the following:

$$\mathcal{A}u = \mathcal{B}T \quad (1)$$

where

$$\mathcal{A} \equiv -\mathcal{F}_M - \rho_m \omega^2 \quad (2)$$

is a mechanical force operator acting on $u(z,x)$ that includes the plate's restoring and damping forces (both captured in $\mathcal{F}_M$) as well as the inertia term. Here $\rho_m$ is the surface mass density in [kg/m$^2$] and the operator $\mathcal{F}_M$ is given, for a thin elastic plate, by[17]

$$\mathcal{F}_M u = -D\nabla^4 u - i\omega\lambda u \tag{3}$$

where $D$ and $\lambda$ are the bending stiffness and damping coefficient, respectively. The operator $\mathcal{B}$ represents the Lorentz force per unit area acting on $T$, and is given by (see Appendix A)

$$\mathcal{B} \equiv B_0 \frac{\partial}{\partial z}. \tag{4}$$

The circuit equation that governs the evolution of the surface current, in the presence of motional electromotive force (EMF) and the applied field, can be written as follows:

$$\mathcal{C}T + \mathcal{D}u = -i\omega B_{app} \tag{5}$$

where

$$\mathcal{C} \equiv -\sigma_s^{-1}\nabla^2 + i\omega K* \tag{6}$$

is the impedance operator and

$$\mathcal{D} \equiv -i\omega B_0 \frac{\partial}{\partial z} \tag{7}$$

is the motional EMF operator. In Eq. (6), the symbol $K*$ indicates convolution with a dipolar field kernel which converts surface dipole density ($T$) into a normal magnetic field on the surface. For an infinitely large plate, $K*$ has a spatial Fourier-domain representation of $(\mu_0/2)k$, where $\mu_0$ is the permeability of vacuum and $k$ is the magnitude of the wavevector (see Appendix A). The two terms of the operator $\mathcal{C}$ correspond to the resistance and self-inductance operators, respectively, because of reasons that will be explained later.

2.1.1. General properties of the coupled equations

Eqs. (1) and (5) constitute inhomogeneous coupled equations for $u$ and $T$, with a source term being proportional to $B_{app}$. Both $u$ and $T$ therefore scale linearly with $B_{app}$. The two operators that are responsible for magneto-mechanical coupling, $\mathcal{B}$ and $\mathcal{D}$, are both proportional the static field $B_0$. In the absence of $B_0$, $u$ and $T$ are decoupled. The frequency dependence appears in the mechanical force operator ($\mathcal{A}$) for damping and inertia, as well as in $\mathcal{C}$ and $\mathcal{D}$ for self- and motion-induced EMF. Finally, if mechanical damping is ignored in Eq. (3), $\mathcal{A}$ is real and no phase shift is involved in the equation of motion Eq. (1). This compares with Eq. (5) where all operators are imaginary except for the resistive one in $\mathcal{C}$. The two terms that single out, namely mechanical damping and electrical resistance, are associated with energy dissipation.

2.1.2. Magnetic damping and stiffening

Magnetic damping and stiffening refer to apparent increase in mechanical damping and stiffness of a conductive object when driven mechanically in the presence of a static magnetic field. While this is not the main topic of this paper, our coupled equations naturally predict existence of such phenomena. Since the analysis lends support to the validity of our equations, discussion of magnetic damping/stiffening is presented in Appendix B.

2.1.3. Motional impedance

The circuit equation analogue of magnetic damping/stiffening is motional impedance (resistance/inductance). This is obtained by the following formal solution of the coupled equations for $T$. We solve Eq. (1) for $u$, and substitute this into Eq. (5) to eliminate $u$. The result is

$$T = (\mathcal{C} + \mathcal{D}\mathcal{A}^{-1}\mathcal{B})^{-1}(-i\omega B_{app}). \tag{8}$$

The magneto-mechanical term $\mathcal{D}\mathcal{A}^{-1}\mathcal{B}$ is proportional to $B_0^2$. We observe that to the extent $\mathcal{C}$ represents the impedance of $T$, the effect of $B_0$ is to add $\mathcal{D}\mathcal{A}^{-1}\mathcal{B}$ to the impedance. We will therefore call this a motional impedance operator $\mathcal{Z}_M$:

$$\mathcal{D}\mathcal{A}^{-1}\mathcal{B} = \mathcal{Z}_M, \tag{9}$$

$$\mathcal{Z}_M \equiv -i\omega B_0 \frac{\partial}{\partial z}(-\mathcal{F}_M - \rho_m\omega^2)^{-1}\left(B_0 \frac{\partial}{\partial z}\right) = -i\omega B_0^2 \frac{\partial}{\partial z}(-\mathcal{F}_M - \rho_m\omega^2)^{-1}\frac{\partial}{\partial z}. \tag{10}$$

This expression, while being formal, can give us considerable insight into the phenomenon of magneto-mechanical resonance. In case where mechanical damping can be ignored, $\mathcal{F}_M$ is real and $\mathcal{Z}_M$ is purely imaginary. The latter can then be expressed as $i\omega\mathcal{L}_M$ where $\mathcal{L}_M$ is motional inductance, a real operator. $i\omega\mathcal{L}_M$ adds to the imaginary (inductive) part of $\mathcal{C}$ in Eq. (8), which is $i\omega K*$. While the conventional inductance operator $K*$ is positive definite for non-zero wavevectors, $\mathcal{L}_M$ is not, and its sign (of the eigenvalues) generally depends on $\omega$. It is therefore possible that the motion-altered inductance of the plate, $(K* + \mathcal{L}_M)$ ceases to be positive and undergoes zero-crossing at a certain frequency for certain current eigenmodes. This is analogous to cancellation of inductive impedance by a capacitor in an LC circuit. We can therefore call this phenomenon magneto-mechanical resonance. At the resonance the total eddy current in the conductive plate inside a static field can be much larger than in zero field. Below we will examine this more closely using plane-wave eigenmodes of an infinitely large plate.

2.1.4. Magneto-mechanical resonance of an infinite plate

Using the definitions for resistance and inductance operators

$$\mathcal{R} \equiv -\sigma_s^{-1}\nabla^2 \tag{11}$$

$$\mathcal{L} \equiv K* \tag{12}$$

we rewrite Eq. (8) as

$$T = (\mathcal{R} + i\omega\mathcal{L} + \mathcal{Z}_M)^{-1}(-i\omega B_{app}). \tag{13}$$

Here $\mathcal{Z}_M = i\omega\mathcal{L}_M$ if one can ignore mechanical damping. As a special case where the plate is infinitely large, the operators $\mathcal{R}, \mathcal{L}, \mathcal{Z}_M$ are simultaneously diagonalized by the plane wave eigenfunctions labeled by the wavevector $\vec{k} = (k_z, k_x)$. The corresponding eigenvalues are:

$$\mathcal{R}: \quad -\sigma_s^{-1}\nabla^2 = \sigma_s^{-1}k^2 \geq 0 \tag{14}$$

$$\mathcal{L}: \quad K* = \frac{\mu_0}{2}k \geq 0 \tag{15}$$

$$\mathcal{Z}_M: i\omega B_0^2 \frac{\partial}{\partial z}(-D\nabla^4 - i\omega\lambda + \rho_m\omega^2)^{-1}\frac{\partial}{\partial z} = \frac{i\omega B_0^2 k_z^2}{Dk^4 - \rho_m\omega^2 + i\omega\lambda}. \tag{16}$$

We used Eq. (3) for $\mathcal{F}_M$ in Eq. (16). At mechanical resonance where $Dk^4 = \rho_m\omega^2$, Eq. (16) avoids singularity by the damping term $i\omega\lambda$ in the denominator. This situation is similar to the mechanical resonance of a simple harmonic oscillator.

Combining Eqs. (13-16), we can write $T$ in the spatial Fourier domain as

$$T(\vec{k}) = \frac{-i\omega}{\sigma_s^{-1}k^2 + i\omega\left(\frac{\mu_0}{2}k + \frac{B_0^2 k_z^2}{Dk^4 - \rho_m\omega^2 + i\omega\lambda}\right)} B_{app}(\vec{k}) \tag{18}$$

where all independent variables are real.

As we increase $\omega$ from zero, we observe that for each spatial mode $\vec{k}$ we encounter two resonances. First is the mechanical resonance where the real part of the denominator of $\mathcal{Z}_M$ vanishes, at $Dk^4 = \rho_m\omega^2$. This is the frequency, to be called $\omega_{mech}$, at which the plate would undergo the maximum displacement in response to pure mechanical excitation, absent Lorentz force. The second, magneto-mechanical resonance occurs when the real part of $\mathcal{Z}_M$ becomes negative and cancels the conventional inductance $(\mu_0/2)k$. It can happen when $Dk^4 < \rho_m\omega^2$, that is, past the mechanical resonance ($\omega > \omega_{mech}$) where the motion and force are out of phase. The importance of mechanical phase reversal in the context of magneto-mechanical resonance is illustrated in Fig. (3).

Below we summarize a few properties of magneto-mechanical resonance of an infinite conducting plate.
 (i) The magneto-mechanical resonance frequency $\omega_{mm}$ satisfies $\omega_{mm} > \omega_{mech}$.
 (ii) $\omega_{mm}$ is a function of the wavevector $\vec{k}$.
 (iii) At the resonance the reactive (inductive) part of the total impedance of the current mode vanishes.
 (iv) To the extent that the impedance of the mode is dominated by the reactive term, magneto-mechanical resonance can greatly increase the eddy current response $T/B_{app}$. However, because of frequency dependence in other terms of Eq. (18), exact maximum of $|T/B_{app}|$ may not occur at $\omega = \omega_{mm}$.

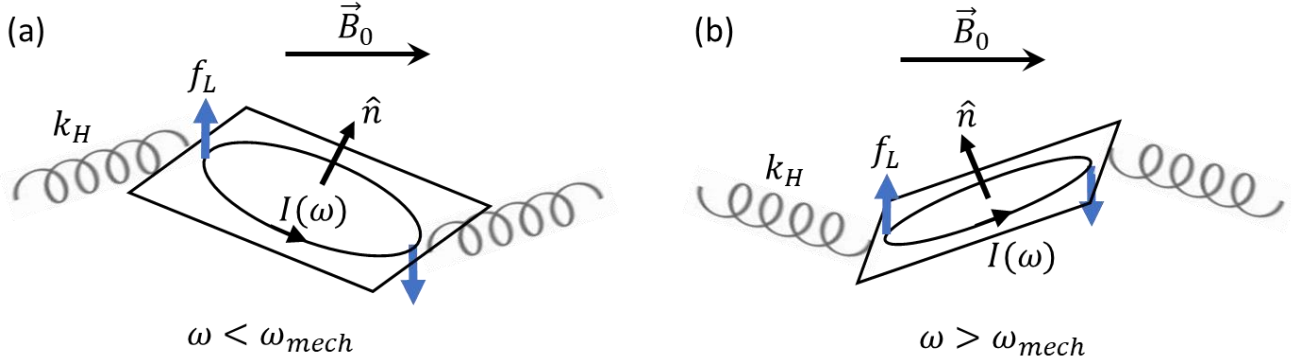

Figure 3. Illustration of motional inductance and onset of magneto-mechanical resonance. (a) At low frequencies ($\omega < \omega_{mech}$), the Lorentz force $f_L$ acting on an oscillating current $I(\omega)$ tilts the plate towards and away from $\vec{B}_0$ in phase with the current. When the current is positive, the tilt of the normal vector $\hat{n}$ is towards $\vec{B}_0$. Such tilt results in positive (locally upward) flux threading the loop, making motional inductance positive. (b) At high frequencies ($\omega > \omega_{mech}$), the Lorentz force and angular tilt of $\hat{n}$ are 180° out of phase; when the current is positive, the tilt is still away from $\vec{B}_0$ trying to catch up. In this case, $\vec{B}_0$ threads the current in the negative direction, or locally downwards, and the motional inductance is negative. If the negative motional inductance cancels the positive self-inductance, the loop is at magneto-mechanical resonance. Then, the current induced by an applied oscillating magnetic field (not shown) is impeded only by the resistance of the loop. $k_H$ denotes the spring constant.

2.1.5. Relationship with eddy image current

From the definition of $\mathcal{Z}_M$ (Eq. 10) as well as the eigenvalue representation in Eq. (18), it can be recognized that as $\omega \to \infty$, $\mathcal{Z}_M \to 0$ and

$$\lim_{\omega \to \infty} T = \lim_{\omega \to \infty} (\mathcal{R} + i\omega\mathcal{L})^{-1}(-i\omega B_{app}) = -\mathcal{L}^{-1}B_{app}. \tag{19}$$

This is the stream function of the eddy image current, to be called $T_{eddyi}$. That is,

$$B_{app} = -\mathcal{L}T_{eddyi}. \tag{20}$$

Physically this means that at high enough frequencies where mechanical motion cannot keep up with the electromagnetic drive, and when inductive impedance dominates resistance, the induced eddy current completely nulls the applied normal magnetic field. We can now express $T$ in terms of $T_{eddyi}$ in Eq. (13),

$$T = (\mathcal{R} + i\omega\mathcal{L} + \mathcal{Z}_M)^{-1}(i\omega\mathcal{L})\, T_{eddyi} \tag{21}$$

while Eq. (18), for an infinite plate, becomes

$$T(\vec{k}) = \frac{i\omega(\mu_0/2)k}{\sigma_s^{-1}k^2 + i\omega\left(\frac{\mu_0}{2}k + \frac{B_0^2 k_z^2}{Dk^4 - \rho_m\omega^2 + i\omega\lambda}\right)} T_{eddyi}(\vec{k}). \tag{22}$$

Eq. (22) shows that, at each frequency and wavevector, the mode amplitude of the shield current differs from an ideal eddy current by a multiplicative factor. The magnitude of this factor can be much larger than unity near $\omega = \omega_{mm}$.

**2.2. Magneto-mechanical coupled equations for a shield plate patterned into an eddy image loop**

Let us assume that we freeze the eddy image current $T_{eddyi}$ in space (Fig. 2b) and discretize it by cutting lines along its contours: $\{T_{eddyi} = c_n\}$ where $c_n$ ($n = 1, 2, \cdots$) is a set of currents that divide $T_{eddyi}$ into $N_{turns} \gg 1$ equidistant levels (Fig. 2c). Following the standard practice of surface coil design[18], we assume that adjacent contours are connected in series and a return path is provided from the last to the first contour so that all the contour lines are traversed in one stroke between the input (P$_1$) and output (P$_2$) terminals. If we now apply a current $i_0$ from P$_1$ to P$_2$ that corresponds to the contour spacing, $i_0 = c_2 - c_1$, such current will approximate the continuous surface current defined by $T_{eddyi}$. The patterned passive shield proposed in this work consists of a closed loop formed by short-circuiting P$_1$ and P$_2$. When acted on by a time-dependent magnetic field from a driving coil, an induced current $I(t)$ in the loop will tend to shield the space opposite to the driving coil from such magnetic field. The goal of this section is to calculate $I(t)$ in the presence of magneto-mechanical coupling.

2.2.1. Current and motional degrees of freedom of the patterned conductor

Let us define a dimensionless static function
$$\tilde{T}_{eddyi}(z,x) \equiv T_{eddyi}(z,x)/i_0. \tag{23}$$
This can be viewed as a continuous (smoothed) version of the "number of turns" function $N(z,x)$ of the discretized surface current described above. $N(z,x)$ is defined by counting (integrating) the number of contour lines (turns) in a direction-sensitive way; that is, by adding $+1$ if the current in the turn crosses the integration path from the left to the right, and $-1$ if in the other direction, as one moves from the edge of the plate to the point (z, x). If the patterned passive shield carries a current $I(t)$ (or $I(\omega)$ in the frequency domain), the corresponding stream function is
$$T(z,x; t \text{ or } \omega) = I(t \text{ or } \omega)N(z,x) \approx I(t \text{ or } \omega)\tilde{T}_{eddyi}(z,x). \tag{24}$$
In the limit $N_{turns} \to \infty$ and $i_0 \to 0$, $N(z,x)$ converges to $\tilde{T}_{eddyi}$. In what follows we will assume such a limit (that is, ignore discretization error) and use $N$ and $\tilde{T}_{eddyi}$ interchangeably.

Importantly, all the surface current degrees of freedom of the conductive plate is now condensed into a single time- or frequency-dependent variable $I$. Note that the scale of $\tilde{T}_{eddyi}$ depends on $i_0$ from Eq. (23), which was somewhat arbitrarily determined during discretization of the eddy image current. This is not a problem because the final stream function $T$ appears as a product of $I$ and $\tilde{T}_{eddyi}$, with opposite $i_0$ dependences.

We will assume that the surface current discretization does not affect the shield plate's mechanical properties. This is realistic if the conductor of the shield plate is backed by a stiff, insulating substrate which dominates the plate's mechanical response.

2.2.2. Equation of motion and circuit equation

The equation of motion for the patterned shield is obtained by substituting Eq. (24) in Eq. (1). The frequency domain equation reads

$$(-\mathcal{F}_M - \rho_m \omega^2) u = B_0 I(\omega) \frac{\partial \tilde{T}_{eddyi}}{\partial z} \tag{25}$$

where the Lorentz force term is now time-space separated and is represented by a single unknown $I(\omega)$.

The circuit equation that governs $I$, coupled to $u$, is obtained from Eq. (5) by (i) replacing $T$ with $I\tilde{T}_{eddyi}$ and (ii) multiplying the resulting equation with $\tilde{T}_{eddyi}$ and integrating over the plane. The left- and right-hand sides of the resulting equation are, respectively,

$$lhs = \left\{ -\sigma_s^{-1} \int \tilde{T}_{eddyi} \nabla^2 \tilde{T}_{eddyi} d^2\vec{r} + i\omega \int \tilde{T}_{eddyi} K * \tilde{T}_{eddyi} d^2\vec{r} \right\} I(\omega) - i\omega B_0 \int \tilde{T}_{eddyi} \frac{\partial u}{\partial z} d^2\vec{r} \tag{26}$$

$$rhs = -i\omega \int \tilde{T}_{eddyi} B_{app} d^2\vec{r}. \tag{27}$$

The two integrals in the parenthesis of Eq. (26) and the integral in Eq. (27) all produce a scalar constant. Specifically, we can identify the following. First,

$$-\sigma_s^{-1} \int \tilde{T}_{eddyi} \nabla^2 \tilde{T}_{eddyi} d^2\vec{r} = R \tag{28}$$

is the conventional DC resistance of the loop where $I$ flows. Second,

$$\int \tilde{T}_{eddyi} K * \tilde{T}_{eddyi} d^2\vec{r} = L \tag{29}$$

is the conventional self-inductance of the same loop. Finally,

$$\int \tilde{T}_{eddyi} B_{app} d^2\vec{r} = \Phi_{app} \tag{30}$$

is the flux coupled to the loop by the applied field $B_{app}$. Eqs. (28-30) are explained in Appendix C.

We can now rewrite Eqs. (26,27) simply as

$$RI + i\omega L I - i\omega B_0 \int \tilde{T}_{eddyi} \frac{\partial u}{\partial z} d^2\vec{r} = -i\omega \Phi_{app}. \tag{31}$$

The third term on the left-hand side can be expressed in terms of $I$ by eliminating $u$ through a formal solution of Eq. (25),

$$u = -B_0 I(\omega)(\mathcal{F}_M + \rho_m \omega^2)^{-1} \frac{\partial}{\partial z} \tilde{T}_{eddyi}. \tag{32}$$

Substituting Eq. (32) to Eq. (31) yields

$$RI + i\omega L I + I \int \tilde{T}_{eddyi} \mathcal{Z}_M \tilde{T}_{eddyi} d^2\vec{r} = -i\omega \Phi_{app} \tag{33}$$

where $\mathcal{Z}_M$ is the motional impedance operator defined in Eq. (10). We can now define the motional impedance parameter of the current loop $\tilde{T}_{eddyi}$ as

$$Z_M = Z_M(\omega) \equiv \int \tilde{T}_{eddyi} \mathcal{Z}_M \tilde{T}_{eddyi} d^2\vec{r}. \tag{34}$$

Recall that if mechanical damping is ignored, $\mathcal{Z}_M$ is purely imaginary and can be written as $\mathcal{Z}_M = i\omega \mathcal{L}_M$. For completeness, we also define the motional inductance of the loop as

$$L_M = L_M(\omega) \equiv \int \tilde{T}_{eddyi} \mathcal{L}_M \tilde{T}_{eddyi} d^2\vec{r}. \tag{35}$$

This signifies the flux per 1 A in the loop that is caused by its physical displacement in the $B_0$ field. The final solution for the current in the presence of magneto-mechanical coupling is

$$I = I(\omega) = \frac{-i\omega \Phi_{app}}{R + i\omega L + Z_M(\omega)}. \tag{36}$$

Magneto-mechanical resonance occurs when the imaginary (reactive) part of $Z_M(\omega)$ becomes negative and cancels the self-inductance $L$. The precise line-shape of $I(\omega)$ depends on the functional form of $Z_M(\omega)$, which is hard to reduce analytically in general. However, $I(\omega)$ can be determined experimentally from the measured impedance of the loop, $Z(\omega) \equiv R + i\omega L + Z_M(\omega)$. To measure $Z$, one should break the short between the terminals P₁ and P₂ and probe their voltage difference while applying sinusoidal current at different $\omega$. Doing this both inside and outside $B_0$ field will allow separating out $Z_M$ from $Z$ of the loop, provided its mechanical properties, including mounting, are kept constant. Such measurements can be useful to characterize a patterned shield regarding its magneto-mechanical behavior.

2.2.3. Relationship with eddy image current

In the high frequency limit, motional impedance vanishes and Eq. (36) converges to $\lim_{\omega \to \infty} I(\omega) = -\Phi_{app}/L$. This limiting current equals the ideal eddy image current in the patterned shield, which was previously defined as $i_0$. That is,

$$\lim_{\omega \to \infty} I(\omega) = -\frac{\Phi_{app}}{L} = i_0. \tag{37}$$

This relation can be used to eliminate $\Phi_{app}$ in Eq.(36):

$$I(\omega) = \frac{i\omega L}{R + i\omega L + Z_M(\omega)} i_0. \tag{38}$$

Multiplying both sides with $\tilde{T}_{eddyi}$, we obtain the patterned shield's current in terms of its stream function:

$$T(z, x; \omega) = I(\omega)\tilde{T}_{eddyi}(z, x) = \frac{i\omega L}{R + i\omega L + Z_M(\omega)} T_{eddyi}(z, x). \tag{39}$$

This highlights the fact that for a patterned shield, the eddy current solution has the same spatial shape as the ideal eddy current, but with an $\omega$-dependent scale factor.

**2.3. Comparison between continuous and patterned passive shields**

It is interesting to compare the surface current solutions for continuous (Eq. (21)) and patterned (Eq. (39)) shields. In the latter, $T$ is a scaled version of the eddy image current, where the proportionality constant corresponds to the operator fraction in Eq. (21) after *all operators are replaced by their expectation values* with respect to the turns-number function $\tilde{T}_{eddyi}$. The expectation value of an operator $\mathcal{O}$ with respect to a function $f(\vec{r})$ is defined as Ref. [19].

$$\langle \mathcal{O} \rangle = \langle f|\mathcal{O}|f \rangle \equiv \int f^*(\vec{r})\mathcal{O}f(\vec{r})d^2\vec{r}. \tag{40}$$

$R$, $L$, $Z_M$, $L_M$ in Eqs. (28,29,34,35) as well as $Z$ considered above correspond to the expectation values of the operators $\mathcal{R}$, $\mathcal{L}$, $\mathcal{Z}_M$, $\mathcal{L}_M$, and $\mathcal{Z}$, respectively. Here we define the total impedance operator $\mathcal{Z} \equiv \mathcal{R} + i\omega\mathcal{L} + \mathcal{Z}_M$.

Further insights can be gained if we consider a special case where the operators $\mathcal{L}$ and $\mathcal{Z}$ can be simultaneously diagonalized. In this case, one can find a set of basis functions $|j\rangle = f_j(z,x), (j = 1,2,\cdots)$ that satisfy

$$\mathcal{L}|j\rangle = \lambda_j |j\rangle \tag{41}$$
$$\mathcal{Z}|j\rangle = \zeta_j |j\rangle \tag{42}$$

where $\lambda_j$ and $\zeta_j$ are the eigenvalues ($\lambda_j$ should not be confused with the damping coefficient $\lambda$). Note that in section 2.1.4, we have considered one such case for an infinitely large plate where $f_j(z,x) \propto \exp(i\vec{k}_j \cdot \vec{r})$. Now consider the following expansion of the eddy image current:

$$T_{eddyi} = \sum_j a_j |j\rangle. \tag{43}$$

In general, both $a_j$ and $|j\rangle$ are complex. We want to express the stream function solutions for the continuous and patterned passive shields using similar expansion.

First, the continuous shield solution can be written, from Eq. (21), as

$$T(continuous) = \mathcal{Z}^{-1}(i\omega\mathcal{L})T_{eddyi} = \sum_j \frac{i\omega\lambda_j}{\zeta_j} a_j |j\rangle. \tag{44}$$

That is, each current mode of $T_{eddyi}$ is multiplied by a factor $i\omega\lambda_j/\zeta_j$. Recall that $\mathcal{Z}$ and therefore its eigenvalues $\zeta_j$ are $\omega$-dependent. Eq. (44) implies that if any of the $\zeta_j$ values (nearly) vanishes at mode-specific magneto-mechanical resonance, the entire solution $T$ undergoes a resonant peak. This partly explains the complexity of magneto-mechanical resonance spectra often encountered in conductive shells and plates[3].

This situation is contrasted with that of a patterned shield. From Eq. (39), we see that

$$T(patterned) = \frac{i\omega\langle\mathcal{L}\rangle}{\langle\mathcal{Z}\rangle} T_{eddyi}. \tag{45}$$

Using the expansion Eq. (43) to compute expectation values, we get

$$T(patterned) = \frac{i\omega \sum_j \lambda_j |a_j|^2}{\sum_j \zeta_j |a_j|^2} \sum_j a_j |j\rangle. \tag{46}$$

Here the singularity of individual modes ($\zeta_j = 0$) does not directly translate into the singularity of $T$ since the denominator is a *weighted sum of all mode eigenvalues*. Zero-crossing of such a sum generally occurs much less frequently than those of individual $\zeta_j$ in aggregate. This can be understood from the fact that if individual $\zeta_j$'s are an *n*-th order polynomial, their weighted sum is still an *n*-th order polynomial, permitting at most *n* zero crossings. One can expect, therefore, that the patterned passive shield will exhibit fewer resonance peaks than the continuous shield. While a practical passive shield is different from an infinite plate, simpler resonance spectrum in a patterned shield was indeed observed in our experiments described below.

## 3. Experimental Methods

### 3.1. Leakage field frequency response function

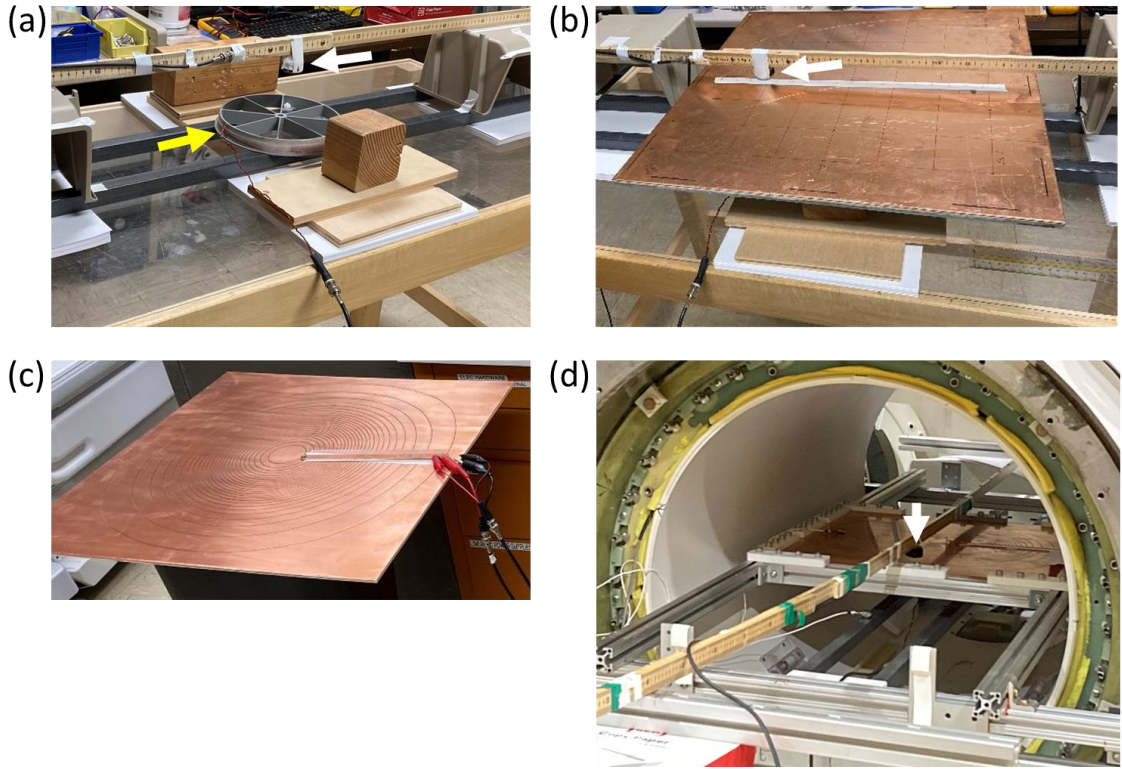

Figure 4. Experimental setup. (a) Driving (yellow arrow) and sensing (white arrow) coils. (b) Solid copper plate placed between the driving (hidden below the plate) and sensing (white arrow) coils. (c) Patterned copper during impedance measurement. The two ends of the spiral pattern were shorted during the FRF measurements. (d) Setup in a whole-body 3T magnet with sensing coil indicated by the white arrow. The copper plate was mounted on an aluminum frame fixed against the magnet end flanges, while the sensing coil was hanging from a wooden pole supported from the floor of the room.

Figure 4 shows the experimental setup. Gradient leakage field was modeled by the magnetic field generated by a 20 cm-diameter, 10-turn circular loop coil ("driving coil") wound from a 1 mm-diameter magnet wire. The field was sensed by a 5 cm-diameter solenoidal pickup coil ("sensing coil") which was mounted on a wooden stick that could be translated horizontally (z direction) above the driving coil. The center-to-center vertical (y direction) distance between the pickup coil and the driving coil was 9.6 cm. The voltage at the pickup coil provided a measure of frequency-weighted magnetic field according to

$$V_{pickup} = -i\omega A_{eff} B_{pickup} \quad (47)$$

where $A_{eff}$ is the effective area of the pickup coil. The effective area was not measured independently, and was left as a common, real-valued scale factor throughout the experiment. This was permissible as we focused on relative changes in the leakage field under different shielding configurations.

The experiment consisted of measuring the frequency response function (FRF), defined as the ratio between $V_{pickup}$ and the driving coil current in the frequency domain, in the following four configurations. (i) First, the FRF was measured at nineteen z locations (z = $-9\Delta z$ to $+9\Delta z$, with $\Delta z$ = 2.54 cm) with no conductive barrier between the two coils. (ii) Second, a 1.6 mm-thick solid copper plate backed by a 9.6 mm-thick fiberglass plate, both measuring 55.9 × 55.9 $cm^2$ in plane, was inserted between the coils (Fig. 4b). The distance between the center of the driving coil and the top surface of the copper plate was 8.1 cm. (iii) Third, the setup (with copper) was moved inside the bore of a 3T whole-body MRI magnet. (iv) Finally, the solid copper plate was replaced by a copper plate with a spiral pattern (Fig. 4c) machined according to the simulated eddy current image of the driving coil (Fig. 4d). The spiral pattern was bridged by a copper strip to make a single closed loop that allowed current to flow in a way to mimic the eddy image.

The eddy image pattern was calculated based on the dimensions of the driving coil and the coil-shield distance in Comsol (Comsol Multiphysics, Burlington, MA, USA), and discretized in Matlab (Mathworks, Natick, MA, USA) into a 25-turn spiral pattern. The pattern was machined on a copper blank plate with 1 mm cut width.

All FRF measurements were carried out with LMS SCADAS Mobile (Siemens PLM Software, Plano, TX, USA) multi-channel data acquisition system (software module: Spectral Testing 11B) with integrated signal generator. A pseudo-white noise voltage signal with 10 kHz bandwidth and 3.125 Hz resolution was fed into an audio amplifier (Crown Audio, Los Angeles, CA, USA), and its output was applied to the driving coil via a 0.1 Ohm current-sensing resistor. The pickup coil voltage and the voltage of the resistor were recorded directly by the data acquisition system, and FRF was computed real-time.

### 3.2. Impedance spectrum

The complex impedance spectrum $Z(\omega)$ of the spiral pattern was measured in the 4[th] (iv) configuration above, with the copper strip electrically open. A dynamic signal analyzer (35670A, Keysight, Santa Rosa, CA, USA) measured the impedance in the frequency range 5 Hz $\leq \omega/2\pi \leq$ 10005 Hz at 1601 logarithmically sampled frequency points. The data was processed in Matlab to calculate theoretical leakage field spectrum of the patterned copper shield in the presence of magneto-mechanical resonance.

## 4. Results

### 4.1. Frequency response function of the leakage field

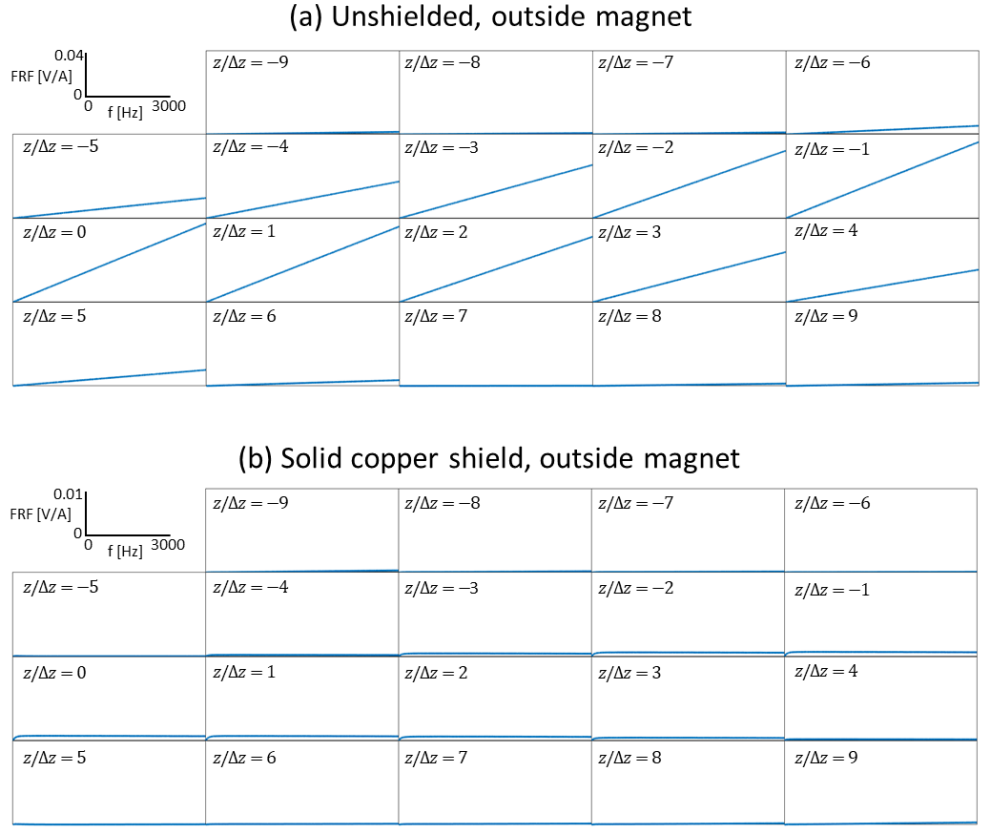

Figure 5. Frequency response functions for unshielded baseline setup (a) and setup with solid copper plate between the driving and sensing coils (b) outside the magnet. Each box contains a current-to-voltage FRF spectrum on the axis defined at the top left corner. 19 FRF's obtained at different z locations of the sensing coil are presented. Note fourfold larger vertical scale in (a). Solid copper outside the magnet provides nearly perfect shielding.

Figure 5 shows the measured FRF at frequencies up to 3 kHz in the first two configurations (with and without copper plate outside the magnet in (a) and (b) respectively). The unshielded FRF (Fig. 5(a)) is linear in frequency as expected from Eq (47). This simply means that there is no frequency-dependent magnetic field in the system without a metallic shield in place. The z-dependence originates from the static (Biot-Savart law) magnetic field profile of the loop coil. Figure 5(b) demonstrates near complete shielding of the driving field by solid copper. Note that the vertical scale is magnified four-fold in Fig. 5(b).

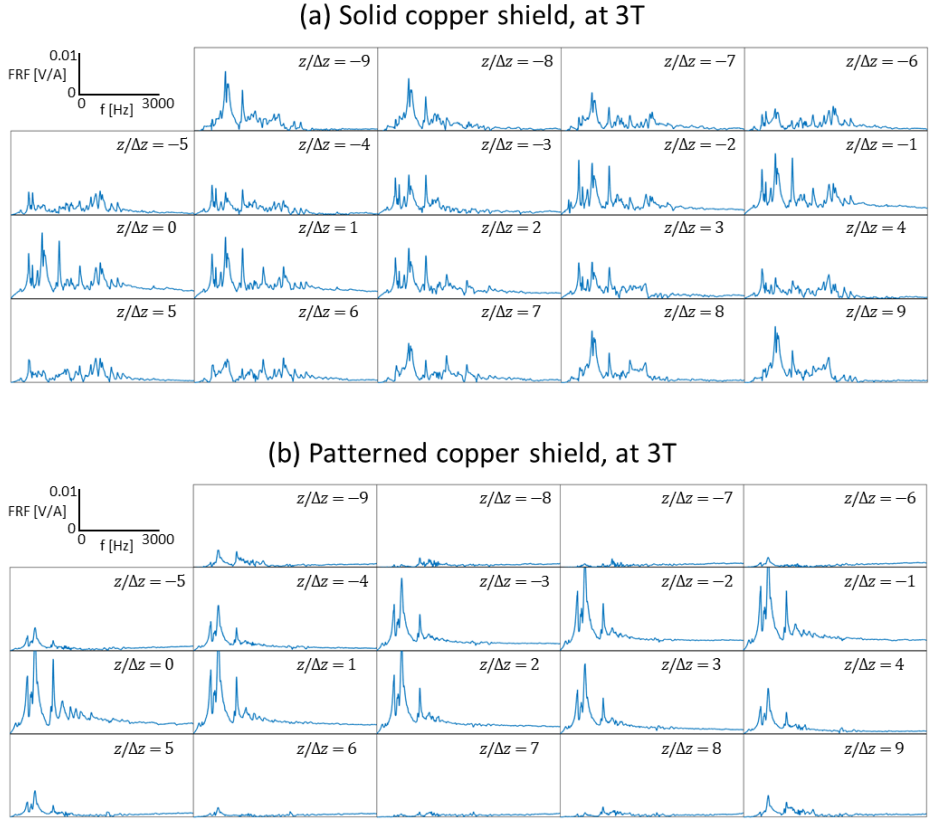

Figure 6. Frequency response functions for solid (a) and patterned (b) copper shields inside a 3T magnet. Compared to Fig. 5(b), shielding is severely disrupted at 3T with numerous magneto-mechanical resonance peaks appearing as a result of motional eddy current. Patterned shield makes the leakage field peaks more localized in space and frequency than solid copper.

Figure 6 shows the measured FRF at 3T with solid and patterned passive copper shields. Comparing Fig. 6(a) with Fig. 5(b), we find dramatic degradation of shielding performance as the copper plate is placed inside a strong static field. This is manifested by many leakage field peaks in FRF of Fig. 6(a) which are caused by magneto-mechanical interaction. Importantly, the FRF spectral shapes are different at different z positions, which indicates multiple surface current modes involved. As expected, the magneto-mechanical response peaks subside in the high frequency limit, above about 2 kHz.

Figure 6(b) contains the main result of this study. It shows that patterning of the copper passive shield drastically changes the spatial and spectral profile of the magnetic field that goes through the shield. In particular, the patterned shield shows substantially reduced leakage field magnitudes in peripheral regions ($|z/\Delta z| > 4$) and at high frequencies ($f > 1$ kHz). Although the leakage field was higher near $z = 0$ and $f < 1$ kHz, Fig. 6 reveals that the patterned-copper FRF is more localized in space and frequency, and exhibits consistent spectral shapes across z. This reflects the fact that the leakage field primarily originates from a single current mode defined by the pattern. Slight change in the spectral patterns near the edge ($|z/\Delta z| > 7$) of the plate is likely caused by localized eddy current in the relatively wide copper traces in the peripheral regions. The spectral and spatial localization of the leakage field can also be appreciated in Fig. 7, where the root-mean-square (RMS) of FRF as a function of z is plotted for different frequency bands. The RMS leakage field greatly increased as the solid copper shield moved into the 3T magnet (black vs red curves in Fig. 7(a,b)), which was then significantly reduced at peripheral z positions (|z| >10 cm) by patterning (blue curves). At high frequencies (1 ~ 3 kHz, Fig. 7b), patterning reduced or held constant the RMS FRF at most z locations (Fig. 7b).

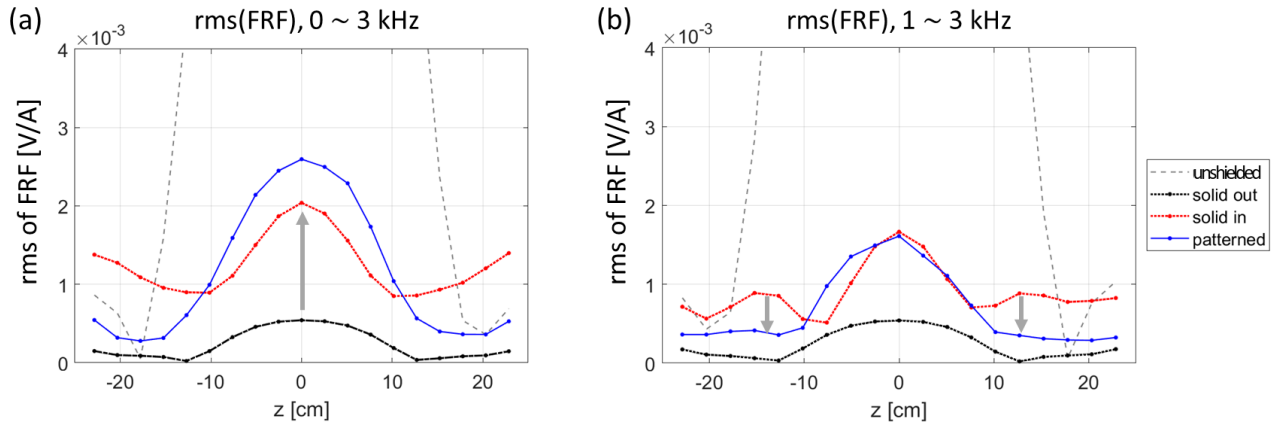

Figure 7. Root-mean-square of the measured pickup-coil FRF over 0 ~ 3 kHz (a) and 1 ~ 3 kHz (b) as a function of z. Large difference between solid copper shields in (red) and out of (black) the magnet highlights leakage field amplification (arrow in (a)) due to motional eddy current. Patterned copper shield in the magnet (blue) makes leakage field more concentrated in frequency and space (arrows in (b)) compared to solid copper.

### 4.2. Impedance spectrum vs FRF

The spectral shape of the patterned-shield leakage field could be well predicted from the impedance measurement as shown in Fig. 8. The real and imaginary parts of the leakage field FRF in Fig. 6(b), at z = 0, are reproduced in Fig. 8(a). All major magneto-mechanical resonance peaks observed in the data are well re-produced in the calculated spectrum shown in Fig. 8(b). This validates (i) that the observed leakage field profiles in Fig. 6(b) are indeed produced by the current in the cut-out pattern, and (ii) that for the patterned loop shield, the magneto-mechanical resonance spectrum can be accurately predicted by impedance measurement of the loop at the same static field. The latter fact significantly simplifies the test of a patterned passive shield as impedance can be easily measured at the terminals of the loop without setting up inductive driving and sensing coils around the shield.

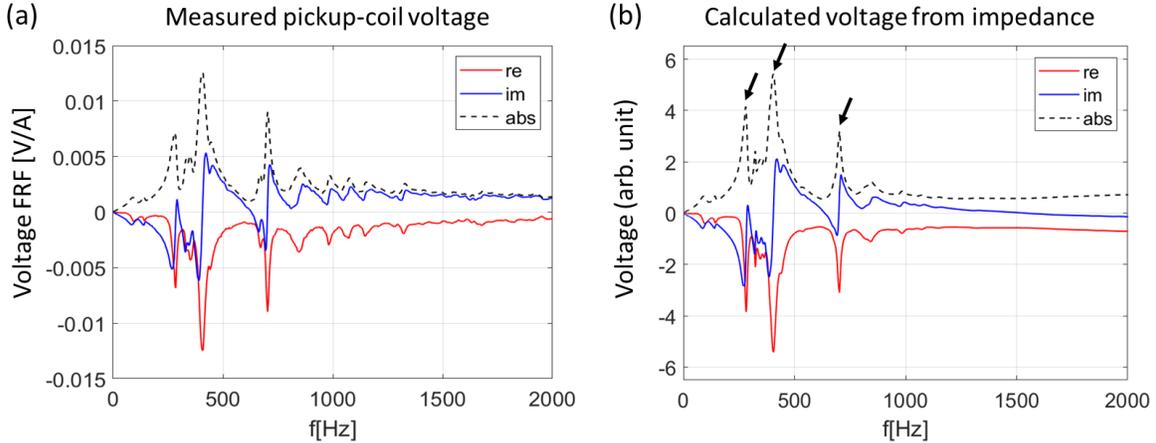

Figure 8. Measured (a) and calculated (b) pickup coil voltage FRF spectra for the patterned copper shield. Measurement was taken with the sensing loop at z = 0. Calculation is based on impedance of the copper spiral pattern open-circuited and measured in the magnet. Major motional eddy current peaks are well reproduced in the calculated spectrum (arrows in (b)).

## 5. Summary and discussion

In this work we developed and analyzed magneto-mechanical coupled equations for a conductive plate in a static magnetic field. We found that coupling between motional and electrical degrees of freedom originates from the Lorentz force acting on moving charges. In the coordinate system chosen, such force is responsible for both the vertical (y-directional) mechanical force acting on x-directional current, as well as the x-directional electromotive force acting on vertically vibrating conductor. These effects manifest themselves as magnetic damping and stiffening in the mechanical response of the plate, and motional inductance and impedance in its electrical response. Our equations showed that, interestingly, magnetic damping is caused by electrical resistance, and the real part of the motional impedance comes from mechanical damping. This is reasonable based on energy conservation; if there is extra damping when an object undergoes forced motion, the dissipated energy has to go somewhere, such as electrical resistive heating. All response coefficients (such as impedance and damping coefficient) that result from magneto-mechanical coupling are proportional to $B_0^2$.

It is interesting and non-trivial that magnetic stiffening is always positive while motional inductance can take both signs. We found that a negative motional inductance can act as a pseudo-capacitance to cancel the electrical inductance of a current mode, leading to resonant amplification of the electrical response of the system. Such magneto-mechanical resonance is expected to occur past a mechanical resonance frequency where the system's mechanical response undergoes phase reversal.

While our analysis was strictly based on a flat conductive plate, it is expected that most qualitative conclusions and insights summarized above will continue to hold in other geometries, such as a cylindrical one which is most relevant in common MRI scanners. This is because the derivation of the main coupled equations, Eqs (1,5), detailed in Appendix A, was drawn from general principles of Newton's and Maxwell's equations, and did not rely in a fundamental way on geometry-specific features and properties. In addition, qualitative observations such as quadratic dependence of magnetic damping on $B_0$ and magneto-mechanical resonance agree with previous, anecdotal reports and analyses based on simpler systems and equations[20][16][21]. To our knowledge, our work is the first to formulate the magneto-mechanical coupling problem on a distributed current as opposed to a lumped-element circuit. Extending the present work to a cylindrical passive shield is left for future investigation.

The main idea of patterned passive shield is to preserve the desired, shielding eddy current while minimizing vibration-induced eddy currents in the presence of a strong static field. This strategy is most effective when the two eddy currents are orthogonal in shape. Our results show that patterned passive shield fundamentally changes the spatial/spectral footprint of the motional eddy current, and can significantly reduce the leakage field in certain frequency bands and spatial regions. Larger leakage field peaks at certain frequency bands may be reduced through structural reinforcements that target specific vibration modes. It should be noted that such measures as well as the exact cut-out patterns needed depend on the particular leakage field of a given driving (gradient) coil.

From Fig. (6), we find that the patterned shield provides shielding efficiency comparable to the continuous conductive plate in the high frequency limit past magneto-mechanical resonances (above ~2 kHz). This validates the idea that high-frequency shielding is provided by one specific current mode, namely the eddy image current, and therefore cutting such a pattern into the continuous plate preserves the plate's shielding capability, above certain frequency thresholds. That one can do this while curtailing other current modes excited by motion is the key idea of the proposed, patterned passive shielding. An obvious limitation of this approach is that a different pattern is needed for each driving coil; in the context of MRI gradient shielding, this means 3 separate passive shielding layers for the 3 axes of the gradient coil assembly. In many cases, however, transverse (X and Y) gradient fields are stronger inducers of magnet heating than the longitudinal one, so two shields may be sufficient to address practical heating issues.

In conclusion, we have developed a theory of magneto-mechanical resonance in distributed current and demonstrated motional eddy current-modified passive shield of an AC magnetic field through mode-limiting cut-outs made on a conductive plate. Our results can help develop methods to effectively control motion-induced eddy current and limit leakage field amplification in a strong static magnetic field. Such methods can be important to minimize magnet-gradient interaction in high-field MRI scanners.

**Data Availability**
The data that support the findings of this study are available from the corresponding author upon reasonable request.

**Author Declaration**
Conflict of interest: The authors are employed by GE HealthCare.

**Acknowledgement**

The authors thank Eric Budesheim for help with mechanical setup of the leakage field experiments. This work was supported in part by the U.S. National Institutes of Health under Grant U01EB026976. The paper does not necessarily represent the views of the funding agency. The authors are employed by and received salary from GE Medical Systems Information Technologies, Inc.
.

# Appendices

## Appendix A. Derivation of the coupled equations

*Equation of motion.* We start from the following general equation of motion governing the mechanical response of the mass element $\rho_m dzdx$ subject to forces,

$$\rho_m \ddot{u} = f_M + f_L. \qquad (A1)$$

This corresponds to the Newton's equation of motion $ma = F$ applied to vertical (y) displacement normalized to the unit area in the zx plane. Here $f_M$ represents the plate's elastic restoring force and damping force, and $f_L$ is the Lorentz force, per area. Soedel (Eq. 4.4.19 and Eq. 8.1.2 of Ref[17]) lists the following expression for the restoring and damping forces for a thin plate:

$$f_M = -D\nabla^4 u - \lambda \dot{u}. \qquad (A2)$$

The Lorentz force under static field $B_0 \hat{z}$ applies to the surface current flowing in the x direction, and can be written as

$$f_L = (\vec{j} \times B_0 \hat{z}) \cdot \hat{y} = -j_x B_0 = -\left(-\frac{\partial T}{\partial z} \cdot B_0\right) = B_0 \frac{\partial T}{\partial z}. \qquad (A2)$$

Combining the above three equations, we get

$$\rho_m \ddot{u} = -D\nabla^4 u - \lambda \dot{u} + B_0 \frac{\partial T}{\partial z} \qquad (A4)$$

which becomes in the frequency domain, after rearranging terms,

$$D\nabla^4 u + i\omega\lambda u - \rho_m \omega^2 u = B_0 \frac{\partial T}{\partial z}. \qquad (A5)$$

This concludes derivation of Eq. (1).

*Circuit equation.* Next we consider the circuit equation governing $T$. We start from the Maxwell's induction equation $\nabla \times \vec{E} = -\partial \vec{B}/\partial t$. Taking a dot product with the surface-normal vector,

$$(\nabla \times \vec{E}) \cdot \hat{n} = -\frac{\partial B_\perp}{\partial t}. \qquad (A6)$$

For a conductive plate with thickness $h$ and bulk conductivity $\sigma$, the E field is related to the surface current density by $\vec{j} = h\vec{J} = \sigma h\vec{E} = \sigma_s \vec{E}$ where $\sigma_s$ is the surface conductivity. Further using the expression $\vec{j} = \nabla \times (T\hat{n})$, we can rewrite the left hand side of the above equation as

$$(\nabla \times \vec{E}) \cdot \hat{n} = \sigma_s^{-1}(\nabla \times \nabla \times (T\hat{n})) \cdot \hat{n} = \sigma_s^{-1}(\nabla(\nabla \cdot (T\hat{n}) - \nabla^2(T\hat{n})) \cdot \hat{n} = -\sigma_s^{-1}\nabla^2 T. \qquad (A7)$$

In the above we used the fact that $T$ does not vary along $\hat{n}$, which leads to zero divergence of $T\hat{n}$ (in Cartesian coordinates), and also that the vector Laplacian amounts to the Laplacian on each Cartesian component. From Eqs. (A6, A7), we obtain what we can call a "surface eddy current equation"

$$\frac{\partial B_\perp}{\partial t} = \sigma_s^{-1}\nabla^2 T. \qquad (A8)$$

Incidentally, this equation can be shown to hold for cylindrical and spherical surface currents as well.

The normal magnetic field $B_\perp$ consists of three parts: applied field $B_{app}$, self-induced field $B_{self}$ that depends on $T$, and motional field $B_{motion}$ that depends on $u$. That is,

$$B_\perp = B_{app} + B_{self}\{T\} + B_{motion}\{u\}. \tag{A9}$$

In light of the dipole density interpretation[22] of $T$, the self-induced field at location $(z, x)$ can be obtained by integrating the dipolar field contributions of $T$ at locations $(z', x')$. Symbolically,

$$B_{self}(z, x, t) = \int\int K(z, x; z', x')T(z', x', t)dz'dx' \equiv K * T(z, x, t) \tag{A10}$$

where we denoted by $K$ the dipolar field kernel that links the surface-normal dipoles to surface-normal dipolar fields. For an infinitely large flat surface, the operator $K *$ is diagonal in the Fourier domain; a plane-wave dipolar density produces a matched plane-wave normal magnetic field with a wavevector-dependent eigenvalue. The mathematical expression for the eigenvalue can be found from Ref[15], whose Eqs (18) and (21) list a $T$ basis function (2D plane wave) and the corresponding normal magnetic field, respectively. After taking the limit of vanishing distance between the dipole and the normal-field planes ($z - z_0 = 0$ in Ref[15]), the ratio between the two equations gives the following eigenvalue,

$$K * \leftrightarrow \frac{\mu_0}{2} k \tag{A11}$$

where $\mu_0$ is the permeability in vacuum and $k = |\vec{k}|$ is the magnitude of the in-plane wavevector.

Finally, the relationship between $u$ and $B_{motion}$ comes from projecting the tilted area element of the displacement field $u(z, x)$, that has non-zero gradient $\nabla u = (\partial u/\partial z, \partial u/\partial x)$, along the $\hat{z}$ direction. The projection equals (details omitted)

$$da_{proj} = -\frac{\partial u}{\partial z} \cdot dzdx. \tag{A12}$$

The negative sign signifies that when $u$ tilts down toward $\hat{z}$, it receives positive (upward) penetration of $B_0$ flux, $d\Phi_{motion} = B_0 \cdot da_{proj}$. By dividing this flux with $dzdx$, we get the motional field

$$B_{motion}\{u\} = -B_0 \frac{\partial u}{\partial z}. \tag{A13}$$

Combining Eqs. (A8, A9, A10, A13), we finally obtain the circuit equation

$$\frac{\partial B_{app}}{\partial t} + K * \frac{\partial T}{\partial t} - B_0 \frac{\partial}{\partial z}\frac{\partial u}{\partial t} = \sigma_s^{-1}\nabla^2 T \tag{A14}$$

which becomes in the frequency domain, after rearranging terms,

$$-\sigma_s^{-1}\nabla^2 T + i\omega K * T - i\omega B_0 \frac{\partial u}{\partial z} = -i\omega B_{app}. \tag{A15}$$

This is Eq. (5).

**Appendix B. Magnetic damping and stiffening**

Suppose our conductive plate is driven by a mechanical force $F_{ext}$ instead of an applied magnetic field. The two coupled equations Eqs.(1) and (5) are modified as

$$\mathcal{A}u = \mathcal{B}T + F_{ext} \tag{A16}$$
$$\mathcal{C}T + \mathcal{D}u = 0. \tag{A17}$$

Equation (A17) can be formally solved for $T$, and substituted in Eq. (A16) to eliminate $T$:

$$\mathcal{A}u = -\mathcal{B}\mathcal{C}^{-1}\mathcal{D}u + F_{ext}. \qquad (A18)$$

Explicitly,

$$(-\mathcal{F}_M - \rho_m \omega^2)u = i\omega B_0^2 \frac{\partial}{\partial z}(-\sigma_s^{-1}\nabla^2 + i\omega K *)^{-1}\frac{\partial}{\partial z}u + F_{ext}. \qquad (A19)$$

While this is not immediately illuminating, its significance can be appreciated when we consider the following two cases.

*Magnetic damping.* If the resistive term in $\mathcal{C}$ dominates the inductive term, for example when $\sigma_s \to 0$ or $\omega \to 0$, Eq. (A19) becomes

$$(-\mathcal{F}_M - \rho_m \omega^2)u = -i\omega B_0^2 \sigma_s \frac{\partial}{\partial z}(\nabla^2)^{-1}\frac{\partial}{\partial z}u + F_{ext}. \qquad (A20)$$

We note that the magneto-mechanical term (first term on the right hand side) is purely imaginary, or 90° shifted from the inertia term. When transferred to the left, it adds to the mechanical damping term $i\omega\lambda u$ contained in $-\mathcal{F}_M u$. We can therefore define a magnetic damping operator

$$\mathcal{G} \equiv B_0^2 \sigma_s \frac{\partial}{\partial z}(\nabla^2)^{-1}\frac{\partial}{\partial z} \qquad (A21)$$

and rewrite Eq. (A20) as

$$(-\mathcal{F}_M + i\omega\mathcal{G} - \rho_m \omega^2)u = F_{ext} \qquad (A22)$$

from which the role of the damping term is clearer. Note that $\mathcal{G}$ is proportional to $B_0^2$ and is positive semi-definite; in the spatial Fourier domain, it reduces to $B_0^2 \sigma_s k_z^2/k^2 \geq 0$. Physically, magnetic damping of a conductive plate results from bending of the plate ($du/dz$) cutting the flux lines of $B_0\hat{z}$, triggering Lorentz force in the form of motional EMF, which in turn causes surface current to flow which then is subject to another Lorentz force acting on the plate in a way to suppress motion. In this process 90° phase shift (time derivative) occurs once between the displacement ($du/dz$) and the motional EMF.

*Magnetic stiffening* If the inductive term dominates $\mathcal{C}$ as in $\sigma_s \to \infty$ or $\omega \to \infty$, the magneto-mechanical term becomes real and Eq. (A19) is approximated as

$$(-\mathcal{F}_M - \rho_m \omega^2)u = B_0^2 \frac{\partial}{\partial z}(K *)^{-1}\frac{\partial}{\partial z}u + F_{ext}. \qquad (A23)$$

The magneto-mechanical term is again proportional to $B_0^2$, and contains $\partial u/\partial z$ corresponding to the plate's bending in the $B_0$ direction. This term opposes the plate's motion in the following sense. Let us define the first term on the right hand side of Eq. (A23) as $-\mathcal{S}u$ where

$$\mathcal{S} \equiv -B_0^2 \frac{\partial}{\partial z}(K *)^{-1}\frac{\partial}{\partial z}. \qquad (A24)$$

It can be shown (see Appendix A) that the Fourier-domain representation of the dipolar field operator $K *$ is $(\mu_0/2)k$ with $k \equiv \sqrt{k_z^2 + k_x^2}$. In this representation, therefore,

$$\mathcal{S} \leftrightarrow B_0^2 \left(\frac{\mu_0}{2}k\right)^{-1} k_z^2 \geq 0. \qquad (A25)$$

This means $\mathcal{S}$ is positive semidefinite and the term $-\mathcal{S}u$ takes a form of restoring force in the Hooke's spring law $F = -k_H x$. Given this consideration, we can call $\mathcal{S}$ magnetic stiffening operator, and re-write Eq. (A23) as

$$(-F_M + S - \rho_m \omega^2)u = F_{ext}. \tag{A26}$$

In practice, both terms in $\mathcal{C}$ are present, and magneto-mechanical suppression of motion has both in-phase and out-of-phase components (with respect to the inertia term). That is, magnetic damping and stiffening co-exist in general, both being proportional to $B_0^2$.

**Appendix C. Derivation of resistance, inductance, magnetic flux expressions**

First, we note that if $B_{app}(z, x)$ is the normal magnetic field on the plate where a surface coil is patterned according to the number of turns function $N(z, x) = \tilde{T}_{eddyi}$, then $\int\int \tilde{T}_{eddyi} B_{app} dz dx$ corresponds to the flux picked up by the coil. This can be understood by dividing $\tilde{T}_{eddyi}$ into a collection of series-connected small solenoidal loops covering the plate whose turn-counts and areas are given by $\tilde{T}_{eddyi}(z, x)$ and $dzdx$, respectively. The total flux in the coil is then given by the integral Eq. (30).

Now if the loop carries a current $I$, the self-generated normal magnetic field is $B_\perp = K * \tilde{T}_{eddyi} I$ by definition of the operator $K*$. The flux picked up by the same loop, divided by $I$, is then given by the integral of Eq. (29). This is by definition the self-inductance of the loop.

Lastly, Eq. (28) can be proven as follows. Over the boundary of the shield plate, the number-of-turns function is zero by definition: $\tilde{T}_{eddyi} = 0$ (plate boundary). Consider a vector field defined in the zx plane, $\vec{V} \equiv \tilde{T}_{eddyi} \nabla \tilde{T}_{eddyi}$. Since this is also zero at the boundary, we can say $\oint \vec{V} \cdot \hat{n} dl = 0$ where the integral is along the boundary and $\hat{n}$ is the unit normal vector at the boundary. From the two-dimensional divergence theorem, this integral equals

$$\oint \vec{V} \cdot \hat{n} dl = \int\int \nabla \cdot \vec{V} dz dx \tag{A27}$$

which expands as

$$\int\int \nabla \cdot \vec{V} dz dx = \int\int \nabla \cdot \left(\tilde{T}_{eddyi} \nabla \tilde{T}_{eddyi}\right) dz dx$$
$$= \int\int \nabla \tilde{T}_{eddyi} \cdot \nabla \tilde{T}_{eddyi} \, dz dx + \int\int \tilde{T}_{eddyi} \nabla^2 \tilde{T}_{eddyi} \, dz dx = 0. \tag{A28}$$

Therefore the left-hand-side of Eq. (28) is

$$-\sigma_s^{-1} \int\int \tilde{T}_{eddyi} \nabla^2 \tilde{T}_{eddyi} dz dx = \sigma_s^{-1} \int\int \nabla \tilde{T}_{eddyi} \cdot \nabla \tilde{T}_{eddyi} \, dz dx. \tag{A29}$$

The fact that this equals the coil's resistance comes from the following consideration. If the coil carries current $I$, the surface current density vector $\vec{j}(z, x)$ is given by

$$\vec{j} = \left(\frac{\partial \tilde{T}_{eddyi}}{\partial x} \hat{z} - \frac{\partial \tilde{T}_{eddyi}}{\partial z} \hat{x}\right) I \tag{A30}$$

from the relation $\vec{j} = \nabla \times (\tilde{T}_{eddyi} I \hat{y})$. This implies that the magnitude the surface current density is related to $\nabla \tilde{T}_{eddyi}$ as

$$|\vec{j}|^2 = I^2 \nabla \tilde{T}_{eddyi} \cdot \nabla \tilde{T}_{eddyi}. \qquad (A31)$$

The total Joule heating of the coil equals the surface integral of the power density, $\sigma_s^{-1}|\vec{j}|^2$, as

$$P = \sigma_s^{-1} \int \int |\vec{j}|^2 dz dx = \sigma_s^{-1} I^2 \int \int \nabla \tilde{T}_{eddyi} \cdot \nabla \tilde{T}_{eddyi}\, dz dx. \qquad (A32)$$

Equating this with $P = I^2 R$, we get Eq. (28).